\RequirePackage{fix-cm}
\documentclass[smallextended]{svjour3}       
\usepackage{graphicx}
\usepackage{fullpage}
\usepackage{lineno,hyperref}
\usepackage{amsmath}
\usepackage{amssymb}
\usepackage{setspace}
\usepackage{subcaption}
\usepackage{float}
\usepackage{graphicx}
\usepackage[numbers]{natbib}
\modulolinenumbers[5]
 
\journalname{Computational Particle Mechanics}
\usepackage[normalem]{ulem}
\usepackage{xcolor, soul}
\sethlcolor{yellow}
\begin{document}

\title{A remedy to mitigate tensile instability in SPH for simulating large deformation and failure of geomaterials}

\author{Tapan Jana$^1$ \and Subhankar Pal$^1$ \and Amit Shaw$^{*1}$ \and L. S. Ramachandra$^1$}
\institute{*Corresponding author \at
              \email{abshaw@civil.iitkgp.ac.in} (Amit Shaw) \\
              \vskip 0cm
           $^1$Department of Civil Engineering, \\
           Indian Institute of Technology Kharagpur, West Bengal, India}

\maketitle

\begin{abstract}
Large deformation analysis in geomechanics plays an important role in understanding the nature of post-failure flows and hazards associated with landslides under different natural calamities. In this study, a SPH framework is proposed for large deformation and failure analysis of geomaterials. An adaptive B-spline kernel function in combination with a pressure zone approach is proposed to counteract the numerical issues associated with tensile instability. The proposed algorithm is validated using a soil cylinder drop problem, and the results are compared with FEM. Finally, the effectiveness of the proposed algorithm in the successful removal of tensile instability and stress noise is demonstrated using the well-studied slope failure simulation of a cohesive soil vertical cut.  
\end{abstract}

\section{Introduction}
Large deformation, landslide hazards, post-failure flow, and stability of geomaterials are the problems of great interest in the field of geotechnical engineering. In the last few decades, there have been considerable efforts to simulate these physical processes primarily via Finite Element Method (FEM) \cite{bathe1975finite, bathe1976elastic, carter1977finite, snitbhan1978elastic, hu1998practical}. To deal with the problems associated with mesh distortion, different approaches of mesh refinement have also been reported in the literature. This includes h-adaptive FE with Super convergent Patch Recovery(SPR) and Remeshing and Interpolation Technique with small strain(RITSS) \cite{hu1998h}, combined rh-adaptive FREM \cite{kardani2013large},  Hessian-based remeshing \cite{cornejo2020combination} etc. Although different adaptive meshing algorithms have been employed, mesh distortion in FEM remains a challenging computational issue, especially in problems in geomechanics which involve severe material flow to the extent that it behaves like a fluid.

In recent years, the smoothed particle hydrodynamics (SPH) \cite{lucy1977numerical, gingold1977smoothed, ShawReid2009, ShawReid2011, ChakrabortyShaw2013, ChakrabortyShaw2014, LahiriShaw2019}, a particle-based (mesh free) method, has emerged as one of the leading computational approaches for assessing geomechanics issues. As SPH does not require any mesh, the problems associated with mesh distortion and the negative value of Jacobian are naturally eliminated. In this regard, one of the pioneering studies was conducted by \citet{bui2008lagrangian}, who simulated the deformation of a vertical cut and post-failure flow of geomaterial using SPH. Therein, the elastoplastic constitutive relation and Drucker-Prager plasticity model were used. \citet{peng2015sph} implemented a hypoplastic constitutive model in the SPH framework to simulate large deformation and slope failure of a cohesionless embankment. \citet{islam2022large} adapted a Total Lagrangian SPH approach to simulate the large deformation and slope failure of cohesive soil. SPH has also been explored in other problems in geomechanics such as liquefaction-induced lateral spreading \cite{naili20052d}, seepage failure analysis \cite{maeda2006development}, soil water interaction \cite{bui2007numerical}, soil cracking modelling \cite{bui2015soil}, soil fragmentation \cite{fan2017parallel}, landslide modelling and run-out analysis (\cite{pastor2014application, liu2021sph}), debris flow or granular flow modelling (\cite{hurley2017continuum}, \cite{bui2006smoothed}, \cite{song2023quantitative}), bearing capacity of shallow foundation \cite{bui2011bearing}.  
 
Despite the potential, SPH has its own shear of computational drawbacks. The significant among them is the tensile instability, which manifests itself in the form of unrealistic clustering and separation of particles, causing inaccurate solutions. The instability was first observed by \citet{schuessler1981comments} in gas dynamics problems and subsequently by \citet{phillips1985numerical} in magnetohydrodynamics problems. Based on one dimensional perturbation analysis, \citet{SWEGLE1995123} studied the stability of SPH and identified the root cause of the instability. \citet{SWEGLE1995123} also derived the condition in which the SPH computation may become unstable. 

The tensile instability has been one of the major thrust areas of the research in SPH. Over the years, several remedial measures have been developed and reported in the literature. 
\citet{guenther1994conservative} presented a conservative smoothing approach that relies on the von Neumann-Richtmyer discrete representation of the conservation of volume. The purpose of this technique was to maintain a steady computing process by eliminating random changes in the field variables. They demonstrated that careful smoothing yields far more stable and precise outcomes than artificial viscosity. This approach was further investigated in \cite{wen1994stabilizing}, \cite{RANDLES1996375}, \cite{HICKS1997209}. After a detailed examination of the conservative smoothing approach, \citet{HICKS2004213} discovered that employing the B-spline finite interpolation method could enhance the stability of 1-dimensional solutions while preserving intricate details.
A novel way for dealing with tensile instability by including stress point in the traditional SPH methodology was addressed in \cite{DYKA1995573, dyka1997stress, randles2000normalized, RABCZUK20041035}. In computing displacement, velocity, and acceleration, the primary particle positions, i.e., actual particle positions (velocity points) are used, whereas the secondary particle positions (stress points) are used to calculate stress, internal energy, and density. While the approach shows better accuracy, tracking, velocity mapping, and subsequent updating of stress points make the approach computationally intensive, especially in higher dimensions. \citet{belytschko2000unified} found that a Lagrangian kernel (based on material coordinates) with stress points is an effective way of ensuring stability in the element-free Galerkin method (EFG). \citet{VIGNJEVIC200067} expanded and enhanced this concept to encompass higher dimensions. \citet{CHALK2020113034} addressed issues related to zero-energy modes and tensile instabilities by incorporating stress points and nodes, along with a new scheme based on fourth order Runge-Kutta for updating stress-point positions. This approach removes the requirement for artificial repulsive forces at the boundary and resolves these challenges without needing to adjust non-physical parameters. 
\citet{morris1996study} highlighted the importance of using kernels with rapidly decreasing Fourier transforms to improve stability in SPH. They showed kernels resembling a Gaussian distribution provide better results, but the computational cost rises with more contributing neighbours. 
To address tensile instability and reduce numerical inaccuracies near the boundaries, \citet{chen1999improvement} devised the corrective smoothed particle method (CSPM) by adding a first order Taylor series expansion to the kernel estimate.  
\citet{SIGALOTTI200823} introduced an alternative solution for mitigating tensile instability in standard SPH fluid simulations. The method employs an Adaptive Density Kernel Estimation (ADKE) algorithm, allowing local adjustments to the kernel interpolant width. This ensures that only the necessary smoothing is applied to the data, optimizing simulation accuracy.
Further, Particle Shifting Techniques (PST), proposed by \citet{XU20096703}, counteract anisotropic particle configurations in incompressible fluid flow by adjusting particle positions. \citet{SUN201725} and \citet{xu2018technique} later extended PST to weakly compressible fluids. 
Background compressive pressure (\cite{morris1997modeling, MARRONE2013456}) is introduced to address instability in low-pressure zones during fluid flow modelling. While effective, setting the background pressure too high can lead to numerical noise, posing a challenge to the method.
Hyperbolic kernel function introduced by \citet{YANG2014199} enables more uniform particle distribution in both two and three-dimensional cases and does not contribute to the unstable increase of stress, allowing for better simulation results.
However, this function has positive second derivatives and can eliminate instability only in compressive regimes.
\citet{monaghan2000sph}, \citet{gray2001sph} developed an artificial stress method, wherein an artificial stress term is introduced in the momentum equation, which produces a short range repulsive force between the neighbouring particles and keeps them in a stable configuration. The artificial stress concept was employed in \cite{bui2008lagrangian, peng2015sph, feng2021large} to study the slope failure and other problems related to geomechanics. The artificial stress technique is by far the most common. However, determining the optimum artificial stress parameters is very elusive. \citet{LAHIRI2020109761} proposed an adaptive kernel in elastic solid problems, where the kernel shape changes with the state of stress to tackle the instability.

In the present study, a pressure-zone based adaptive SPH framework is proposed for simulating large deformation problems in geomechanics. The proposed adaptive algorithm is constituted based on the observations from Swegle's \cite{SWEGLE1995123} stability analysis. In SPH, particles interact with each other through a compactly supported kernel function. The force between any two particles depends on the kernel gradient, which changes with distance between the interacting particles. When a tensile force separates two particles, the inter-particle force must increase with the inter-particle distance and prevent the particles from moving away. However, due to the typical shape of the kernel gradient, the inter particle force builds up until it reaches a limiting value, and then it gradually fades away thereafter. So, after reaching the maximum value, the force between two particles decreases with increased distance between them, resulting in negative stiffness. This manifests in the form of an unphysical separation of particles. In the compression dominated region, unphysical clumping of particles occurs for the same reason. As per Swegle's analysis, this situations arises when $\sigma W'' < 0$ where, $\sigma$ is the stress (compression positive) and $W''$ is the second derivative of the kernel function. The proposed algorithm attempts to continuously adapt the shape of the kernel centred at a given point depending on the average state of stress in its neighbouring zone. This paper is organized as follows. The governing equations and constitutive relations of the soil are discussed in Section \ref{ConstiRelation}. Section \ref{SPHequations} discusses the equations and their discretized version associated with the problem. The proposed adaptive algorithm is presented in Section \ref{Adpative}. Validation of the developed SPH framework is demonstrated in Section \ref{SPHvalidation}, and the large deformation analysis of soil is shown in Section \ref{SlopeFailure}. Conclusions are drawn in Section \ref{Conclusion}. 
  
\section{Governing equations and constitutive model} \label{ConstiRelation}
\subsection{Conservation Equation}
The motion of any continuum body is governed by the conservation equations (mass, momentum, and energy), which can be written in Lagrangian form using Einstein's indicial notations as,
\begin{equation}
	\frac{d\rho}{dt} = -\rho\frac{\partial v^{\beta}}{\partial x^{\beta}}
	\label{density}
\end{equation}
\begin{equation}
	\frac{dv^{\alpha}}{dt} = \frac{1}{\rho}\frac{\partial\sigma^{\alpha\beta}}{\partial x^{\beta}} 
	\label{moment}
\end{equation}
\begin{equation}
	\frac{de}{dt} = \frac{\sigma^{\alpha\beta}}{\rho}\frac{\partial v^{\beta}}{\partial x^{\beta}} 
	\label{energy}
\end{equation}
where $\rho$ stands for the density; $x^\beta$, $v^\alpha$ and $\sigma^{\alpha \beta}$ represent the components of the position vector, velocity vector, and Cauchy stress tensor respectively; $e$ denotes the specific energy; $\alpha$ and $\beta$ are the spatial coordinates.  
The Cauchy stress tensor can further be decomposed in terms of pressure and deviatoric stress as $\sigma^{\alpha\beta}=-p\delta^{\alpha\beta}+\tau^{\alpha\beta}$ where $p$ is pressure term calculated as $p=-\frac{1}{3}\sigma^{\alpha \alpha}$ and $\delta^{\alpha\beta}$ represents the Kronecker delta.
\subsection{Constitutive relation}
Total strain rate tensor is divided into elastic and plastic parts as, 
\begin{equation} \label{eq:1}
	\dot\varepsilon^{\alpha\beta}=\frac{1}{2}\left(\frac{\partial v^\alpha}{\partial x^\beta}+\frac{\partial v^\beta}{\partial x^\alpha}\right)=\dot\varepsilon^{\alpha\beta}_e +\dot\varepsilon^{\alpha\beta}_p.
\end{equation}
The elastic part of the strain rate may be expressed using generalized Hooke's law as, 
\begin{equation} \label{eq:2}
	\dot\varepsilon^{\alpha\beta}_e=\frac{1+\nu}{E}\dot\sigma^{\alpha\beta}-\frac{\nu}{E}\dot\sigma^{kk}\delta^{\alpha\beta}
\end{equation}
where, $E$ and $\nu$ are the modulus of elasticity and Poisson's ratio respectively; and $k$ is dummy index. Using plastic potential $Q$ and the non-associative flow rule, the plastic strain rate may be expressed as 
\begin{equation} \label{eq:3}
	\dot\varepsilon^{\alpha\beta}_p=\dot\Lambda\frac{\partial Q}{\partial\sigma^{\alpha\beta}}
\end{equation}
where $\Lambda$ is the plastic multiplier. Substituting Equations \ref{eq:2} and \ref{eq:3} into Equation \ref{eq:1}, the total strain rate may be written as,
\begin{equation} \label{eq:4}
	\dot\varepsilon^{\alpha\beta}=\frac{1+\nu}{E}\dot\sigma^{\alpha\beta}-\frac{\nu}{E}\dot\sigma^{kk}\delta^{\alpha\beta}+\dot\Lambda\frac{\partial Q}{\partial\sigma^{\alpha\beta}}
\end{equation}
Now, Equation \ref{eq:4} can be rearranged to obtain the total stress rate tensor as 
\begin{equation} \label{eq:5}
	\dot\sigma^{\alpha\beta}=2G\dot\varepsilon^{\alpha\beta}+\frac{3K\nu}{1+\nu}\left[\dot\varepsilon^{kk}-\dot\Lambda(\frac{\partial Q}{\partial \sigma^{pq}}\delta^{pq})\right]\delta^{\alpha\beta}-2G\dot\Lambda\frac{\partial Q}{\partial \sigma^{\alpha\beta}}
\end{equation}
where shear modulus $G$ and bulk modulus $K$ is defined as
\begin{equation} \label{eq:6}
	G=\frac{E}{2(1+\nu)}  ;    K=\frac{E}{3(1-2\nu)}
\end{equation}
After some algebraic manipulation, Equation \ref{eq:5} can be further simplified as 
\begin{equation} \label{eq:7}
	\dot\sigma^{\alpha\beta}=2G\dot e^{\alpha\beta}+K\dot\varepsilon^{kk}\delta^{\alpha\beta}-\dot\Lambda\left[\left(K-\frac{2G}{3}\right)\frac{\partial Q}{\partial \sigma^{pq}}\delta^{pq}\delta^{\alpha\beta}+2G\frac{\partial Q}{\partial \sigma^{\alpha\beta}}\right]
\end{equation}
\subsection{Drucker-Prager plasticity model} \label{DP_model}
In Drucker-Prager yield criteria, yield surface $F$ depends on both deviatoric as well as hydrostatic component of stress tensor and may be defined as,
\begin{equation}  \label{eq:8}
	F\left(p,J_2\right)=\sqrt{J_2}-\alpha_{\phi}p-k_c
\end{equation}
where, $J_2=\frac{1}{2}\tau^{\alpha\beta}\tau^{\alpha\beta}$ is the second invariant of deviatoric stress tensor, and $\alpha_{\phi}$ and $k_c$ are related to cohesion $c$ and angle of internal friction $\phi$ in the following manner
\begin{equation}  \label{eq:9}
	\alpha_{\phi}=\frac{3tan\phi}{\sqrt{9+12tan^2\phi}} ;  k_c=\frac{3c}{\sqrt{9+12tan^2\phi}}
\end{equation}
The plastic potential differs from the yield function for the non-associative flow rule and may be expressed as, 
\begin{equation}  \label{eq:10}
	Q\left(p,J_2\right)=\sqrt{J_2}-\alpha_{\varphi}p
\end{equation}
where, $\alpha_\varphi$ is related to the dilatancy angle $\varphi$ as,
\begin{equation}  \label{eq:11}
	\alpha_{\varphi}=\frac{3tan\varphi}{\sqrt{9+12tan^2\varphi}}. 
\end{equation}
Finally, the total stress rate $\dot\sigma^{\alpha\beta}$ and the rate of plastic multiplier $\dot\Lambda$ may be obtained as,
\begin{equation} \label{eq:12}
	\dot\sigma^{\alpha\beta}=2G\dot e^{\alpha\beta}+K\dot\varepsilon^{kk}\delta^{\alpha\beta}-\dot\Lambda\left[K\alpha_\varphi\delta^{\alpha\beta}+\frac{G}{\sqrt{J_2}}\tau^{\alpha\beta}\right]
\end{equation}
\begin{equation}  \label{eq:13}
	\dot\Lambda=\frac{\displaystyle{\left(\frac{G}{\sqrt{J_2}}\right)}\tau^{\alpha\beta}\dot\varepsilon^{\alpha\beta}+\alpha_{\phi}K\varepsilon^{kk}}{G+K\alpha_{\phi}\alpha_{\varphi}}
\end{equation}
To ensure material frame indifference, Jaumann stress rate is employed by including the spin terms, and accordingly, Equation \ref{eq:12} may be rewritten as,
\begin{equation} \label{eq:jaumann}
	\dot\sigma^{\alpha\beta}=\sigma^{\alpha k}\dot\omega^{\beta k} +\sigma^{k \beta}\dot\omega^{\alpha k} +2G\dot e^{\alpha\beta}+K\dot\varepsilon^{kk}\delta^{\alpha\beta}-\dot\Lambda\left[K\alpha_\varphi\delta^{\alpha\beta}+\frac{G}{\sqrt{J_2}}\tau^{\alpha\beta}\right]
\end{equation}
where $\dot\omega^{\alpha\beta}$ denotes components of spin tensor and may be obtained as, 
\begin{equation} \label{eq:omega}
	\dot\omega^{\alpha\beta}=\frac{1}{2}\left(\frac{\partial v^\alpha}{\partial x^\beta}-\frac{\partial v^\beta}{\partial x^\alpha}\right)
\end{equation}
\subsection{Stress scaling back}
\begin{figure}[htp]
	\centering
	\includegraphics[width=10cm]{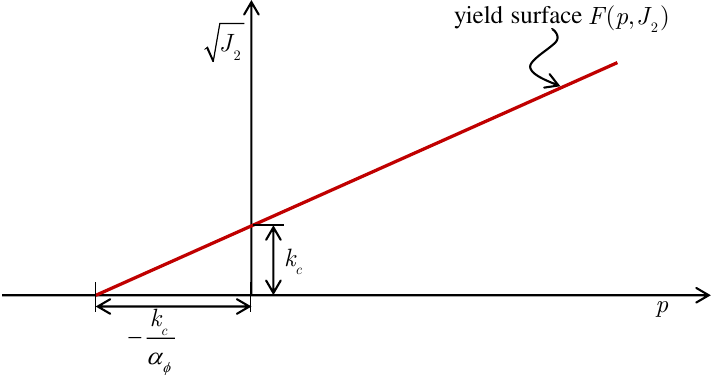}
	\caption{yield surface on $p-\sqrt{J_2}$ plane}
	\label{fig:plasticity}
\end{figure}

General representation of the yield function on $p - \sqrt{J_2}$ plane is presented in Figure \ref{fig:plasticity}. During numerical simulation, the state of stress may exceed the yield surface. In such situations, the state of stress (pressure and deviatoric stress both) is brought back onto the yield surface.
\subsubsection{Pressure correction}
For $p<-\dfrac{k_c}{\alpha_{\phi}}$, the stresses are scaled back using the following equation to ensure $p\ge \displaystyle{-\frac{k_c}{\alpha_{\phi}}}$.
\begin{equation} \label{eq:15}
	\widehat\sigma^{\alpha\beta}=\sigma^{\alpha\beta}+\left(p+\frac{k_c}{\alpha_\phi}\right)\delta^{\alpha\beta}
\end{equation}
\subsubsection{deviatoric stress correction}
Yield surface is defined by the equation $F=0$ or $\sqrt{J_2}=\alpha_{\phi}p+k_c$. If the state of stress exceeds the yield surface, i.e., $\sqrt{J_2}>\alpha_{\phi}p+k_c$, to scale back the deviatoric stresses on to the yield surface, a scaling parameter is defined as 
\begin{equation} \label{eq:17}
	\xi=\frac{\alpha_{\phi}p+k_c}{\sqrt{J_2}} 
\end{equation}
and subsequently, the deviatoric stresses are multiplied with the scaling factor, keeping hydrostatic components unchanged as,
\begin{equation} \label{eq:18}
	\widehat\sigma^{\alpha\beta}=-p\delta^{\alpha\beta}+\xi\tau^{\alpha\beta}
\end{equation}
\section{SPH discretization} \label{SPHequations}
In SPH, the computational domain is represented by a set of particles, which interact with each other such that the conservation equations and constitutive relations described in Section \ref{ConstiRelation} are satisfied at every particle. Let $\bar{\Omega}=\partial\Omega\bigcup\Omega$ be the computational domain and is discretized into a set of particles positioned at ${\{\mathbf{x}_k}\}_{k=1}^N$, where $N$ is the total number of particles. Let ${\{m_k}\}_{k=1}^N$, ${\{\rho_k}\}_{k=1}^N$, ${\{p_k}\}_{k=1}^N$,${\{v^{\alpha}_k}\}_{k=1}^N$, ${\{{\sigma_k}^{\alpha \beta}}\}_{k=1}^N$ be the discrete values of mass, density, pressure, velocity and Cauchy stress respectively at the particles. Continuous representation of the discrete values of the field variables is obtained through a kernel approximation as $<f(\mathbf{x})> \cong \sum_{j \in \mathbb{N}} f_j W(\mathbf{x}-\mathbf{x}_j,\kappa h)\frac{m_j}{\rho_j}$, where $W\left(\mathbf{x},\kappa h\right)$ is a bell-shaped compactly supported kernel function with centre at $\mathbf{x}$ and smoothing length $\kappa h$, $\kappa$ being the support size of the kernel in its parametric space; and $\mathbb{N}$ is the particles in influence domain defined as $\mathbb{N} = \lbrace j \in \mathbb{Z}^+ ~|~ |\mathbf{x}-\mathbf{x}_j| < \kappa h \rbrace$. 

Substituting the kernel representation of the field variables in Equations \ref{density}-\ref{energy}, the conservation equations, in their discrete form, may be obtained as 

\begin{equation}
	\frac{d\rho_i}{dt} = \sum_{j \in \mathbb{N}^i} {m_j ({v_i}^\beta-{v_j}^\beta)W_{ij,\beta}}
	\label{disdensity}
\end{equation}
\begin{equation}
	\frac{dv_i^{\alpha}}{dt} = \sum_{j \in \mathbb{N}^i} {m_j\left(\frac{\sigma_i^{\alpha\beta}}{\rho_i^{2}}+\frac{\sigma_j^{\alpha\beta}}{\rho_j^{2}}-\Pi_{ij}\delta^{\alpha \beta}\right) W_{ij,\beta}}
	\label{dismoment}
\end{equation}
\begin{equation}
	\frac{de_i}{dt} = -\frac{1}{2} \sum_{j \in \mathbb{N}^i} {m_j({v_i}^\beta-{v_j}^\beta)\left(\frac{\sigma_i^{\alpha\beta}}{\rho_i^{2}}+\frac{\sigma_j^{\alpha\beta}}{\rho_j^{2}} - \Pi_{ij} \delta^{\alpha \beta}\right) W_{ij,\beta}}
	\label{disenergy}
\end{equation}
where, $W_{ij,\beta} = \frac{\partial W(x_i-x_j, \kappa h)}{\partial x_i^{\beta}}$ is the kernel gradient and  $\mathbb{N}^i = \lbrace j \in \mathbb{Z}^+ ~|~ |\mathbf{x}_i-\mathbf{x}_j| < \kappa h \rbrace$ is the influence domain of $i$-th particle (i.e., the indices of the particles which interact with $i$-th particle). In Equations \ref{dismoment} and \ref{disenergy}, \textbf{$\Pi_{ij}$} is the artificial viscosity, which is used to stabilize the numerical simulation in the presence of shock or any sudden jump in the field variables \cite{monaghan1988introduction}. The following form of artificial viscosity \cite{monaghan1989problem} is used in the present study. 
\begin{equation}
	\Pi_{ij}=
	\begin{cases}
		\frac{-\gamma_1\bar{c}_{ij}\mu_{ij} + \gamma_2\mu^2_{ij}}{\bar{\rho}_{ij}}  & \text{for} \, \boldsymbol{v}_{ij}.\boldsymbol{x}_{ij} < 0,\\
		0 & \text{otherwise}, \\
	\end{cases}
	\label{artviscosity}
\end{equation}
where, $\mu_{ij}= \frac{h(\boldsymbol{v}_{ij}.\boldsymbol{x}_{ij})}{|\boldsymbol{r}_{ij}|^2 + \epsilon h^2}$,  $\bar{c}_{ij} = \frac{c_i + c_j}{2}$, $\bar{\rho}_{ij} = \frac{\rho_i + \rho_j}{2}$, and $\epsilon$ is taken as $0.01$ to prevent singularity when $r_{ij}$ becomes very small. $\gamma_1$ and $\gamma_2$ are the parameters through which the strength of artificial viscosity can be regulated. 
 
\begin{figure}[htp]  
	\centering
	\includegraphics[width=0.45\textwidth,height=1.75in]{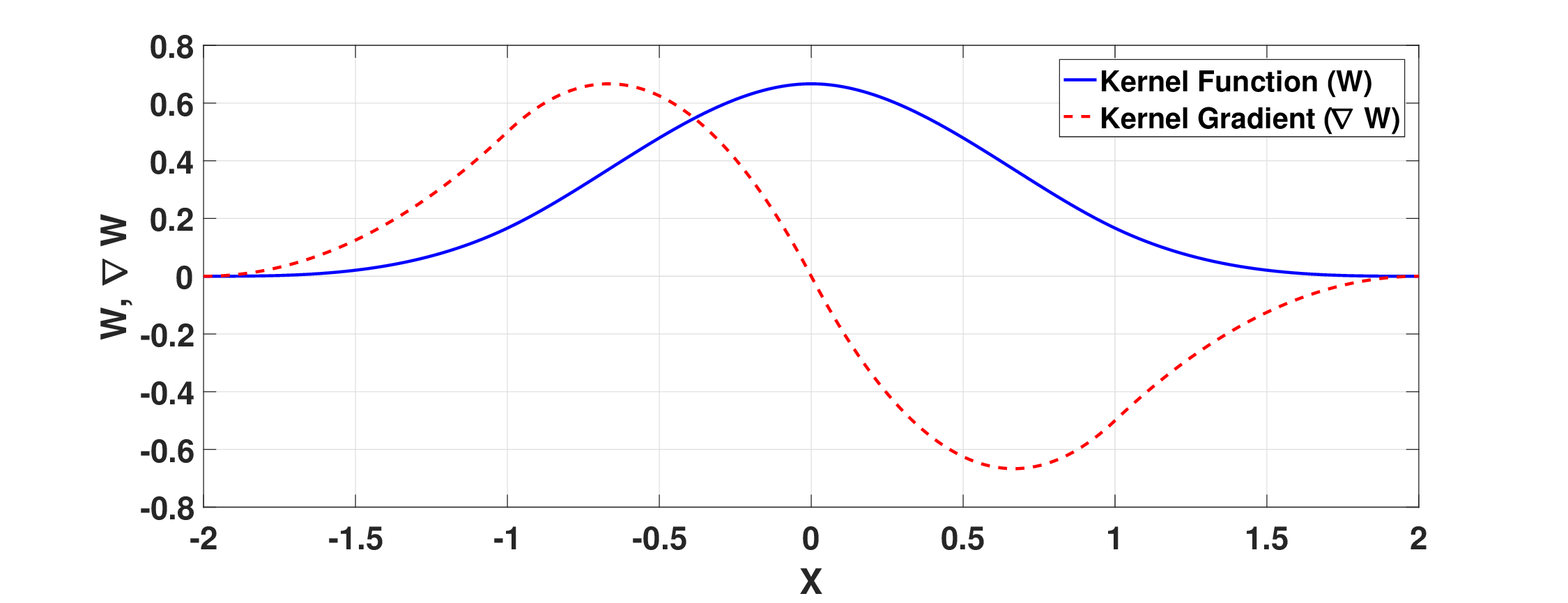}
	\caption{Cubic B-spline kernel function in 1-D}
	\label{KernelFunction}
\end{figure}
Strain rate (Equation \ref{eq:1}) and rotation rate (Equation \ref{eq:omega}) equations may be discretized as, 
\begin{equation}
	\varepsilon^{\alpha \beta}_{i}=\frac{1}{2}\sum_{j \in \mathbb{N}}\frac{m_j}{\rho_j}\left[\left(v^{\alpha}_j-v^{\alpha}_i\right)W_{ij,\beta}+\left(v^{\beta}_j-v^{\beta}_i\right)W_{ij,\alpha} \right]
	\label{eq:strainrate_disc}
\end{equation}
\begin{equation}
	\omega^{\alpha \beta}_{i}=\frac{1}{2}\sum_{j \in \mathbb{N}}\frac{m_j}{\rho_j}\left[\left(v^{\alpha}_j-v^{\alpha}_i\right)W_{ij,\beta}-\left(v^{\beta}_j-v^{\beta}_i\right)W_{ij,\alpha} \right]
	\label{eq:rotationrate_disc}
\end{equation}
Finally, the discretized form of stress rate corresponding to non associative flow rule may be expressed as,
\begin{equation} \label{eq:jaumann_disc}
	\dot\sigma^{\alpha\beta}_i=\sigma^{\alpha k}_i\dot\omega^{\beta k}_i +\sigma^{k \beta}_i\dot\omega^{\alpha k}_i +2G\dot e^{\alpha\beta}_i+K\dot\varepsilon^{kk}_i\delta^{\alpha\beta}-\dot\Lambda_i\left[K\alpha_\varphi\delta^{\alpha\beta}+\frac{G}{\sqrt{J_2}}\tau^{\alpha\beta}_i\right]
\end{equation}
where the plastic multiplier rate $\dot\Lambda_i$ is defined as, 
\begin{equation}  \label{eq:lambda_disc}
	\dot\Lambda_i=\frac{\displaystyle{\left(\frac{G}{\sqrt{J_2}}\right)}\tau^{\alpha\beta}_i\dot\varepsilon^{\alpha\beta}_i+\alpha_{\phi}K\varepsilon^{kk}_i}{G+K\alpha_{\phi}\alpha_{\varphi}}
\end{equation}
To estimate the unknown variables from the discretized governing equations, the conventional predictor-corrector time stepping approach \cite{monaghan1989problem} is used in the present study. The required time-step is obtained through CFL (Courant-Fredrich-Levy) condition.
\section{Pressure zone based adaptive algorithm} \label{Adpative}
Swegle's stability criteria \cite{SWEGLE1995123} provides an insight into the genesis of the tensile instability. For two particles $i$ and $j$ in tension (inter-particle distance increases), their interaction becomes unstable if $ W^{''}_{ij} \geq 0$. If the interacting particles are in compression (inter-particle distance decreases), the condition for instability becomes $W^{''}_{ij} \leq 0$. In the present study, the proposed methodology is an attempt to translate these conditions into an algorithm where the shape of the kernel function is adapted depending on the state of stress such that the particles always interact in a stable manner. To this end, the B-spline basis function defined over a symmetric but non-uniform knot vector is used as the kernel function. The shape of the kernel is modified by changing the position of the knots. A concept of a pressure zone is introduced to characterize the state of stress at a particle in relation to its neighbouring particles. The methodology is described in this section.  
 
\subsection{B-spline basis function}
B-spline basis functions are piece-wise polynomials that may be constructed following the recurrence relation developed by deBoor, Cox, and Mansfield \cite{piegl1996nurbs}. The $I^{th}$ B-Spline basis function of $P^{th}$ degree is formulated as,
\begin{equation}
	\label{B-spline basis}
	\begin{split}
		&N_{I,0}=\begin{cases}
			1, & \text{if $\zeta_{I}\leq \zeta<\zeta_{I+1}$},\\
			0, & \text{otherwise}.
		\end{cases} \\
		N_{I,P}(\zeta)&=\frac{\zeta-\zeta_{I}}{\zeta_{I+P}-\zeta_I}N_{I,P-1}(\zeta)+\frac{\zeta_{I+P+1}-\zeta}{\zeta_{I+P+1}-\zeta_{I+1}}N_{I+1,P-1}(\zeta)  
	\end{split}
\end{equation}
where $\Xi=\lbrace\zeta_1,\zeta_2, \zeta_3,...,\zeta_{m} | \zeta_{I} \in \mathbb{R} \rbrace$ is the non decreasing sequence of real numbers coined as the knot vector with $\zeta_{I}$ being the position of the $I$-th knot. The $I$-th basis function $N_{I,P}$ is defined over $[\zeta_I,\zeta_{I+1}...,\zeta_{I+P+1}]$. Its shape within the support $[\zeta_I,\zeta_{I+P+1}]$ may be modified by changing the positions of the intermediate knots $\lbrace \zeta_{I+1},...\zeta_{I+P} \rbrace$ and the support can be changed by changing the positions of the extreme knots. 

Following the above relation, $N_{0,3}$ constructed over $\Xi=\{-b,-a,0,a,b\}$ gives a symmetric cubic B-spline basis which is considered as the kernel in the present formulation. After normalization, the expression of the kernel function may be obtained as,    
\begin{equation}
	\label{kernel}
	\begin{split}
		W(q,h)=\alpha_c \begin{cases}
			\frac{(a+b)q^3-3abq^2+a^2b^2}{a^2b(a+b)}, & \text{if $0\leq q<a$}\\
			\frac{(b-q)^3}{b(b^2-a^2)}, & \text{if $a\leq q<b$}\\
			0, & \text{if $b\leq q$}
		\end{cases}     
	\end{split}
\end{equation}
where $\alpha_c=\frac{2}{bh}$ for 1D and $\frac{10(a+b)}{\pi b(a^2+ab+b^2)h^2}$ for 2D, determined by using the kernel's normalization condition, $\int_\Omega W(\textbf{x}-\textbf{x$'$},h)d\textbf{x$'$}=1$. In this construction, there is only one intermediate knot positioned at $a$, and by changing the value of $a \in (0,b)$, the shape of the kernel can be changed as illustrated in Figure \ref{fig:kernel_and_deri}.

\subsection{Adaptive Algorithm: Concept}
\begin{figure}[htp]
	\centering
	\includegraphics[width=0.6\textwidth]{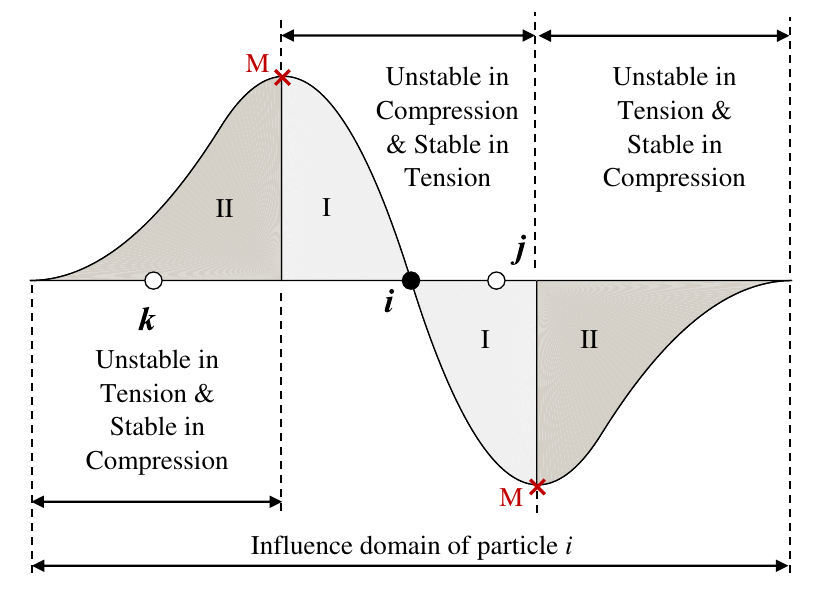}
	\caption{Stable zone in particle interaction}
	\label{fig:zone}
\end{figure}

When a particle interacts with other particles within its influence domain, the stability of the interaction depends on how the neighbouring particles are positioned on the graph of the kernel of the given particle and also on the state of stress at the particle with respect to its neighbours as depicted in Figure \ref{fig:zone}. Herein, the interaction of particle $i$ with particle $j$ becomes unstable in compression, and with particle $k$ becomes unstable in tension. Now, the essence of the algorithm is to modify the shape of the kernel such that the neighbouring particles are covered by the stable zone of the modified kernel. The change in shape may be realized by changing the position of the intermediate knot $a$ in the Cubic B-spline given by Equation \ref{kernel}. 

\begin{figure}
	\begin{subfigure}[b]{0.49\textwidth}
		\centering
		\includegraphics[width=\textwidth]{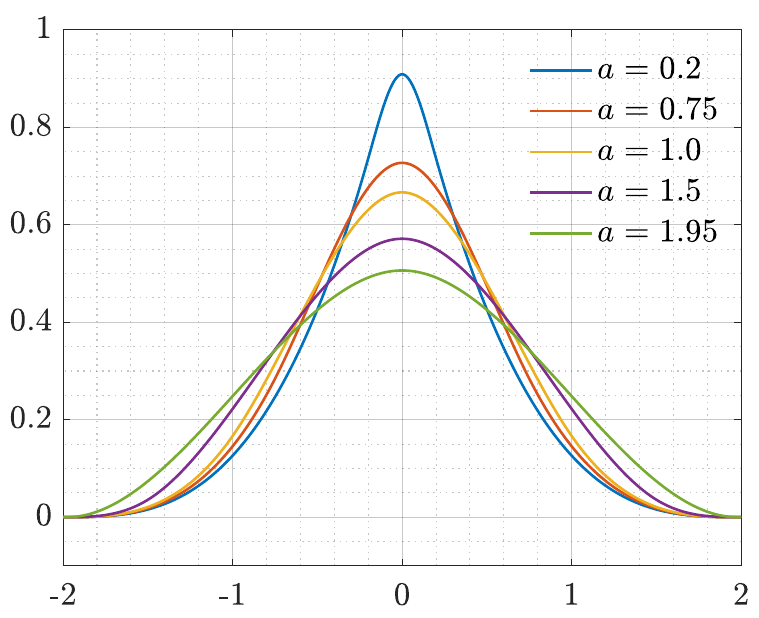}
		\subcaption{$W$ }
		\label{fig:kernel_different_a}
	\end{subfigure}
	\hfill
	\begin{subfigure}[b]{0.49\textwidth}
		\centering
		\includegraphics[width=\textwidth]{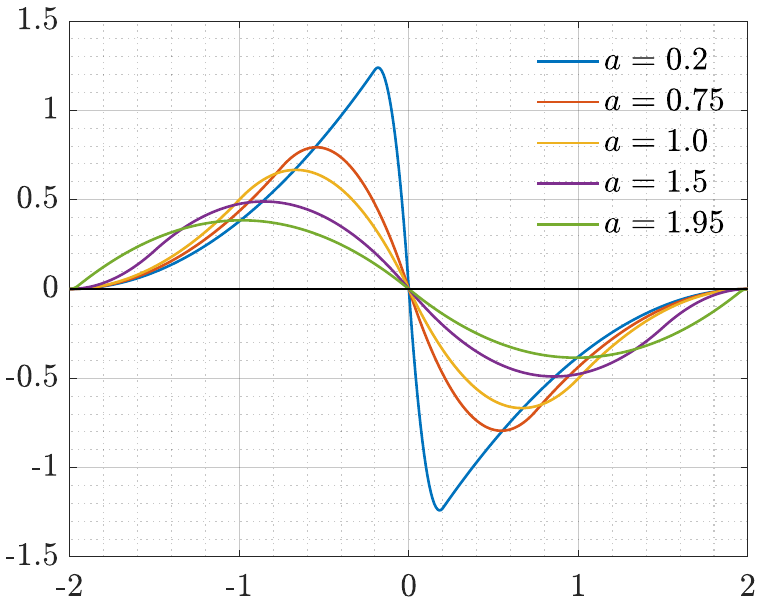}
		\caption{$W^{\prime}$ }
		\label{fig:kernel_deri_different_a}
	\end{subfigure}
	\caption{The kernel $W$ and gradient of kernel $W^{\prime}$ for different values of $a$ and keeping $b$ as 2}
	\label{fig:kernel_and_deri}
\end{figure}

The algorithm requires the determination of the direction and amount by which  the extremum of the kernel function gradient is to be shifted to maintain Swegle's stability criteria. For a cubic B-spline kernel function with smoothing length $h$, $W^{'}$ attains its peak value at a distance $\frac{ab}{a+b}h$ from the center position (i.e., $x=0$). Now if the peak of the kernel gradient is to be positioned at a distance $r$ from the centre, the value of $a$ needs to be taken as $a=\frac{br}{bh-r}$. The change in the extremum of $W'$ with $a$ is illustrated in Figure \ref{fig:kernel_and_deri}. From Figures \ref{fig:zone} and \ref{fig:kernel_and_deri} it is evident that the particles in compression (approaching) $a$ need to be shifted towards centre, which results in increasing zone II (i.e., stable in compression) and the particles in tension (moving away) $a$ need to be shifted towards the boundary of the influence domain resulting in increasing zone I (i.e., stable in tension). While doing such exercise, it is to be ensured that the interacting particles lie within their newly formed respective stable zone. This constitutes the philosophy of the adaptive algorithm. 

Translation of the above concept into a computational code faces two major hurdles. First, due to the compact support of the kernel, it may not be possible to bring all the particles in the influence domain into a common stable zone. Second, particle  $i$  may experience different nature of interaction (i.e., tension or compression) with its different neighbours $j\in \mathbb{N}^i$. To deal with this, in the proposed algorithm, the shape of the kernel is modified such that its peak (point M in Figure \ref{fig:zone}) is shifted to an extent to bring the immediate neighbours into the stable zone. This works because the immediate neighbours have the major contribution and if their interactions are made stable, the overall stability of the particle interaction may be ensured. Moreover, instead of treating every interaction of particle $i$ with its neighbour $j\in \mathbb{N}^i$ separately, a pressure-zone concept is introduced to characterize the overall nature of stress at $i$ with respect to its neighbours, and that is used as the guiding parameter to decide whether $a$ is to be increased or decreased. This is discussed in the following sub-section.

\subsection{Pressure zone based criteria}
\begin{figure}[htp]
	\centering
	\includegraphics[width=.6\textwidth]{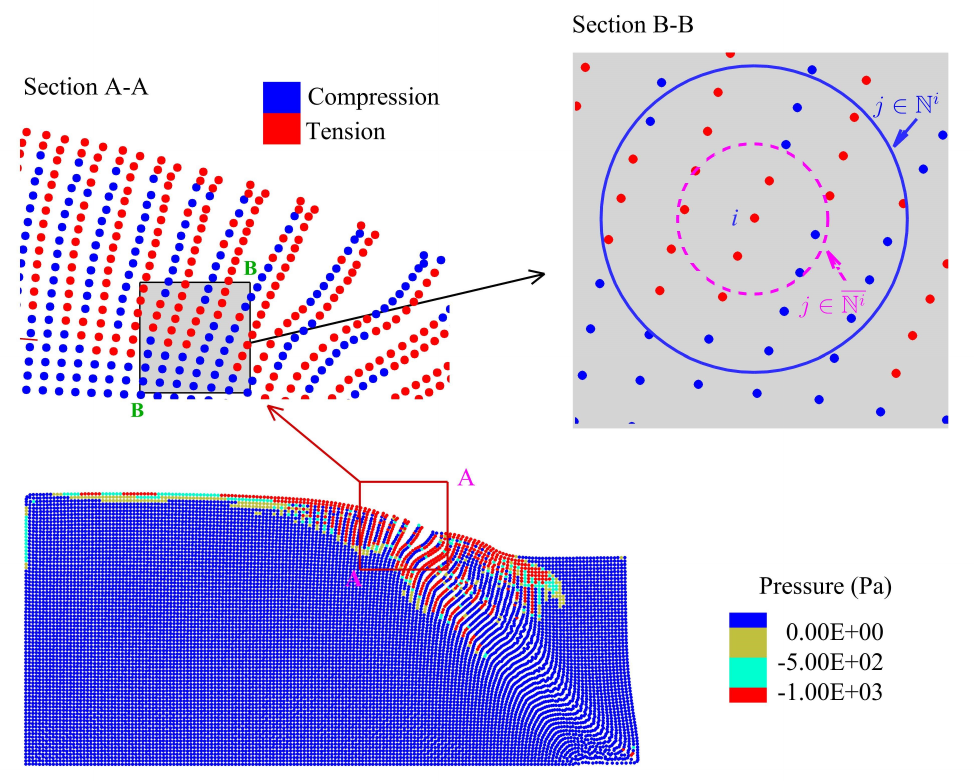}
	\caption{Deformed configuration along with pressure contour at $t$=0.8 s using conventional SPH and demonstration of pressure zone}
	\label{fig:pressure_zone_1}
\end{figure}
While the slope failure analysis of cohesive soil is taken up in detail in Section \ref{SlopeFailure}, for better comprehension of the adopted pressure zone based criteria, the deformed configuration of the soil mass at $0.8$s obtained using the standard SPH (shown in Figure \ref{fig:pressure_zone_1}) is used herein. From Section A-A, it may be observed that the tensile instability is manifested in the form of repetitive compression and tension layers of soil particles. Section B-B of Figure \ref{fig:pressure_zone_1} shows the influence domain ($\mathbb{N}^i$) of a given particle $i$. The immediate neighbour $\overline{\mathbb{N}^i} \subseteq \mathbb{N}^i$ is also shown in the same figure. 

The instability may arise both in tension and compression. However, the literature suggests that the instability in tension is more prominent and can severely pollute the solution \citep{morris1997modeling, monaghan2000sph, gray2001sph}. In fact, most of the existing remedial measures are directed towards removing the instability in tension. Therefore, if any particle encounters immediate neighbours with different natures of stress (i.e., some neighbours are in tension and some are in compression as in the case of particle $i$ in Figure \ref{fig:pressure_zone_1}), then there is a possibility of growth of instability due to tension in the interaction between particle $i$ and $\lbrace j \in \overline{\mathbb{N}^i} ~|~ p(j)<0 \rbrace$ depending on where the particle $j$ lies (zone I or zone II in Figure \ref{fig:zone}). In such situation, $a$ is increased to increase the stable zone in tension (i.e., zone I) such that it covers all the immediate neighbours. The value of $a$ is determined as $a=\frac{br_i}{bh-r_i}$, where $r_i$ is the distance of the furthest immediate neighbour of $i$-th particle. The maximum value that $a$ can take is the cut off of the B-spline kernel, i.e., $b$ and this occurs when $r_i = bh/2$. For $r_i > bh/2$, $a$ is not further increased and is kept same as $b$. The procedure for determining the furthest immediate neighbour is discussed in Section \ref{furthest}.

\begin{figure}[htp]
	\centering
	\includegraphics[width=1\textwidth]{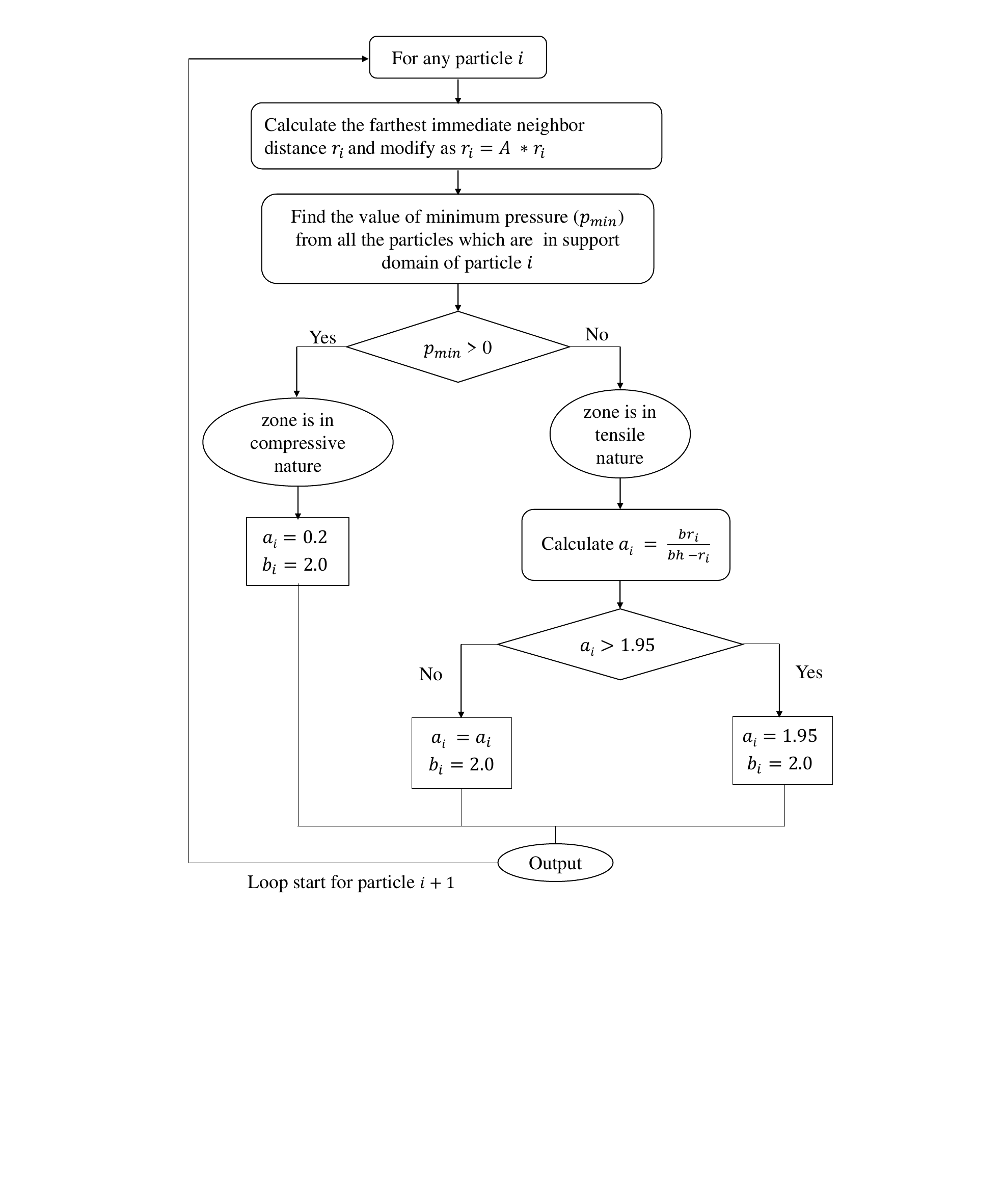}
	\caption{Flowchart for the adaptive algorithm of particle $i$}
	\label{Flowchart}
\end{figure}

If all the immediate neighbours of $i$ are in compression, i.e., $p(j) > 0 ~\forall j \in \overline{\mathbb{N}^i}$, the value of $a$ needs to be decreased such that closest immediate neighbour goes outside of the extremum of $W'$. This ensures that all the immediate neighbour particles under compression are in the stable zone (i.e., zone II). In such situation, it is observed that any small value of $a$, say 0.2, will suffice. The flowchart outlining the different steps involved in the adaptive algorithm is shown in Figure \ref{Flowchart}.
 
\subsection{Estimation of the farthest immediate neighbour} \label{furthest}
The adaptive algorithm technique to remove tensile instability relies heavily on estimating the distance to the furthest immediate neighbour.
\citet{randles2000normalized} presented a technique for locating closest neighbours. They performed an inverse mapping of points in the candidate list and estimated the convex hull in the mapped space to get the nearest neighbours. In the present study, the value of $r_i$ is determined using a strain-based approach.
Consider a two-dimensional (2D) grid of particles, originally separated by a distance $\Delta X^0$ along the X-direction and a distance $\Delta Y^0$ along the Y-direction. If $\dot{\epsilon}_{i,t}^{xx}$, $\dot{\epsilon}_{i,t}^{yy}$ and $\dot{\epsilon}_{i,t}^{xy}$ are the components of strain rate at particle $i$ at $t$-th time step, then the line segment $\Delta X^0$ changes its length to $\Delta X^t_i = \Delta X^0 + \sum\dot{\epsilon}_{i,t}^{xx} \Delta X^{t-1} dt$ and the relative displacement of one end of the line segment with respect to the other is $\sum\dot{\epsilon}_{i,t}^{xy} \Delta X^{t-1} dt$. Similarly, the length of the line segment $\Delta Y^0$ changes to $\Delta Y^t_i = \Delta Y^0 + \sum\dot{\epsilon}_{i,t}^{yy} \Delta Y^{t-1} dt$, and the relative displacement of one end of the line segment with respect to the other is $\sum\dot{\epsilon}_{i,t}^{yx} \Delta Y^{t-1} dt$. 
The aforementioned strain rates can be calculated as
\begin{equation}
	\label{Estimate_ri_2}
	\begin{split}
		&\dot{\epsilon}_{i,t}^{xx} =\sum_j \frac{m_j}{\rho^{t-1}_j}(u^{t-1}_j-u^{t-1}_i)\frac{\partial W_{ij}}{\partial x_i}, \\
		&\dot{\epsilon}_{i,t}^{yy} =\sum_j \frac{m_j}{\rho^{t-1}_j}(v^{t-1}_j-v^{t-1}_i)\frac{\partial W_{ij}}{\partial y_i}, \\
		&\dot{\epsilon}_{i,t}^{xy} =\frac{1}{2} \sum_j \frac{m_j}{\rho^{t-1}_j} \left[ (v^{t-1}_j-v^{t-1}_i)\frac{\partial W_{ij}}{\partial x_i} + (u^{t-1}_j-u^{t-1}_i)\frac{\partial W_{ij}}{\partial y_i} \right].
	\end{split}
\end{equation}
Now, the rectangle of sides $\Delta X^0$ and $\Delta Y^0$ becomes a rhombus whose diagonals ($S_1$ and $S_2$) may be determined as, 
\begin{equation}
	\label{Estimate_ri_1}
	\begin{split}
		&S_1 = \sqrt{(\Delta X^t _i + \sum\dot{\epsilon}_{i,t}^{xy} \Delta Y^{t-1} dt)^2 + (\Delta Y^t_i + \sum\dot{\epsilon}_{i,t}^{yx} \Delta X^{t-1} dt)^2}\\
		&S_2 = \sqrt{(\Delta X^t_i  - \sum\dot{\epsilon}_{i,t}^{xy} \Delta Y^{t-1} dt)^2 + (\Delta Y^t_i - \sum\dot{\epsilon}_{i,t}^{yx} \Delta X^{t-1} dt)^2}.
	\end{split}
\end{equation}
The \emph{farthest immediate neighbour} of the $i$-th particle at time step $t$ is considered to be at a distance $\mathrm{max}(S_1, S_2)$ away. 

\section{Validation of the SPH framework for elasto-plastic analysis of cohesive soil} \label{SPHvalidation}
\begin{figure}[H]  
	\centering
	\includegraphics[width=9cm]{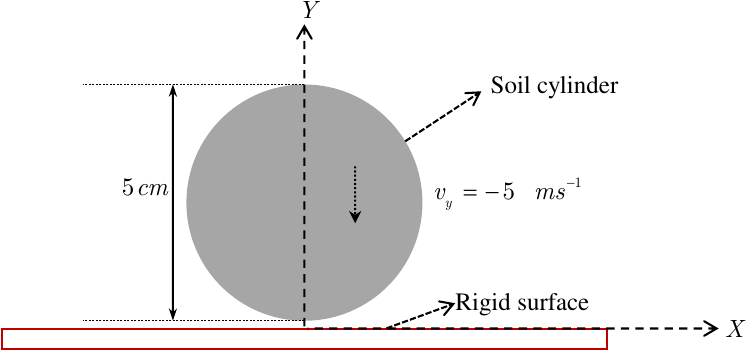}
	\caption{initial configuration}
	\label{fig:drop1}
\end{figure}

The issue of tensile instability in the context of slope-failure analysis of cohesive soil is discussed in the next section. Therein, it also demonstrates that how the tensile instability can be controlled via the proposed algorithm. However, before that, the computational framework for elasto-plastic analysis of cohesive soil is validated in this section. To this end, a 2D plane strain problem of a soil cylinder with a diameter of 5 cm and impacting on a rigid surface with a velocity of 5 m/s is considered. The initial configuration is shown in Figure \ref{fig:drop1}. The initial gap between the cylinder and the rigid surface is kept as 0.15 cm. The effect of gravity is also considered in the simulation.
The soil medium is modelled as an elastoplastic material with Drucker-Prager yield criteria and a non-associative flow rule with a zero dilatancy angle. The pressure and deviatoric stress corrections are applied as discussed in Section \ref{DP_model}. Total 8037 numbers of particles with inter-particle spacing of 0.0005 m are used to discretize the cylindrical soil specimen, and the rigid surface is constructed using 2005 number of particles. The material properties and other computational data are provided in Tables \ref{tab:drop_table_1} and \ref{tab:drop_table_2}, respectively. 
\begin{table}[H]
	\centering 
	\caption{Material properties of soil}
	\setlength{\tabcolsep}{5pt}  
	\renewcommand{\arraystretch}{1.5}  
	\begin{tabular}{c  c  c  c  c }
		\hline
		$\rho_0$ &  $E$ & Poisson's ratio$(\nu)$ & Cohesion $(c)$ & Internal friction angle$(\phi$)\\
		\hline
		1850 kg/m$^3$ & 5 MPa & 0.2 & 30 kPa & $22^\circ$ 
	\end{tabular}
	\label{tab:drop_table_1}
\end{table}
\begin{table}[H]
	\centering 
	\caption{SPH Computational data}
	\setlength{\tabcolsep}{5pt}  
	\renewcommand{\arraystretch}{1.5}  
	\begin{tabular}{c  c  c  c  c }
		\hline
		Particle spacing $(s)$ & Smoothing length $(h)$ & $\Delta t$  & \multicolumn{2}{c}{Artificial viscosity coefficients}\\
		\cline{4-5}
		&&& $\alpha$ &$\beta$ \\
		\hline
		0.0005 m & 0.00075 m & $10^{-6} $s & 1.0 & 1.0
	\end{tabular}
	\label{tab:drop_table_2}
\end{table}
\begin{figure}[htp]
	\centering
	\includegraphics[width=0.5\textwidth]{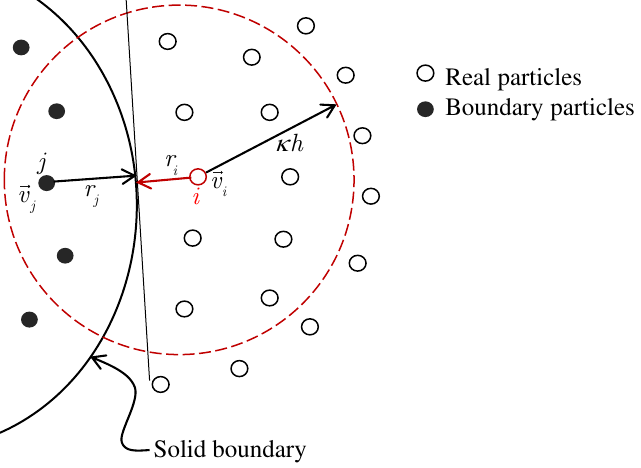}
	\caption{Boundary treatment}
	\label{fig:block_fig_2}
\end{figure}
To simulate the solid surface, the nonslip boundary condition proposed by \citet{morris1997modeling} is implemented. Three layers of boundary particles with spacing same as the initial particle spacing are used. To ensure zero normal as well as tangential velocity along the boundary line, if a boundary particle $j$ comes into the influence domain of a real particle $i$, a fictitious velocity $\displaystyle \mathbf{v}_{j}=-\zeta\mathbf{v}_{i}$ is imposed on the boundary particle $j$ where $\displaystyle\zeta=\frac{r_j}{r_i}$, with $r_i$ and $r_j$ being the distance of the solid boundary from the real and fictitious particle respectively as illustrated in Figure \ref{fig:block_fig_2}. It is to be noted that the fictitious velocity is only used in the conservation equations and strain rate calculation, and the positions of boundary particles are not updated as the boundary is considered to be stationary. Finally, the velocity difference between the real particle $i$ and boundary particle $j$ is determined as,  
\begin{equation}\label{eq:relative_vel}
	\mathbf{v}_{ij}=\mathbf{v}_{i}-\mathbf{v}_{j}
	=\chi\mathbf{v}_{i} 
\end{equation}
where $\displaystyle\chi=min(\chi_{max},1+\zeta)$ and the value of $\chi_{max}$ is generally taken as 1.5 - 2 to eliminate extremely high velocity of boundary particle.

\begin{figure}[H]
	\centering
	\captionsetup[subfigure]{labelformat=empty}
	\begin{subfigure}[b]{0.75\textwidth}
		\centering
		\includegraphics[width=\textwidth]{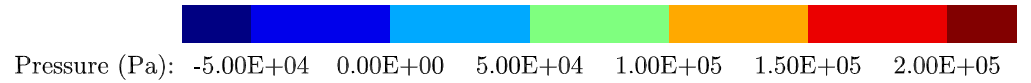}
		\label{fig:contour_drop}
	\end{subfigure}
	\hfill
	\begin{subfigure}[b]{0.40\textwidth}
		\centering
		\includegraphics[width=\textwidth]{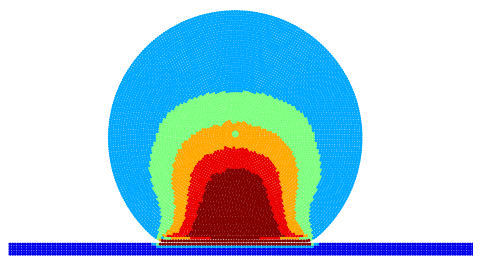}
		\caption{(a)}
		\label{fig:drop_1ms}
	\end{subfigure}
	\hfill
	\begin{subfigure}[b]{0.40\textwidth}
		\centering
		\includegraphics[width=\textwidth]{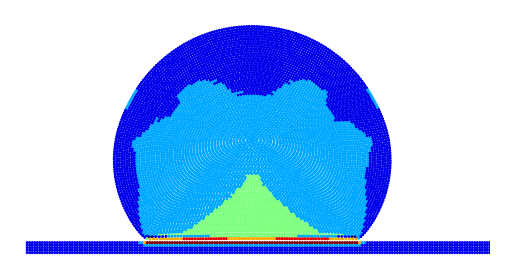}
		\caption{(b)}
		\label{fig:drop_2ms}
	\end{subfigure}
	\hfill
	\begin{subfigure}[b]{0.40\textwidth}
		\centering
		\includegraphics[width=\textwidth]{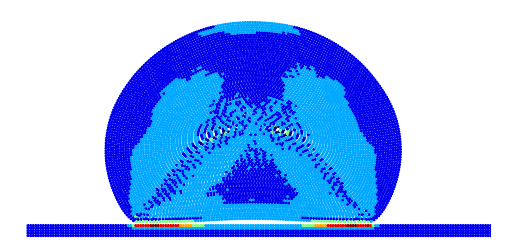}
		\caption{(c)}
		\label{fig:drop_3ms}
	\end{subfigure}
	\hfill
	\begin{subfigure}[b]{0.40\textwidth}
		\centering
		\includegraphics[width=\textwidth]{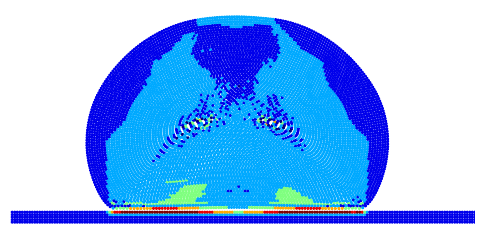}
		\caption{ (d)}
		\label{fig:drop_4ms}
	\end{subfigure}
	
	\caption{Deformation pattern after impact (a) $t$= 1 ms, (b) $t$= 2 ms, (c) $t$= 3 ms, (d) $t$= 4 ms}
	\label{fig:soil_drop_snapshot}
\end{figure}

\begin{figure}[H]
	\centering
	\includegraphics[width=9cm]{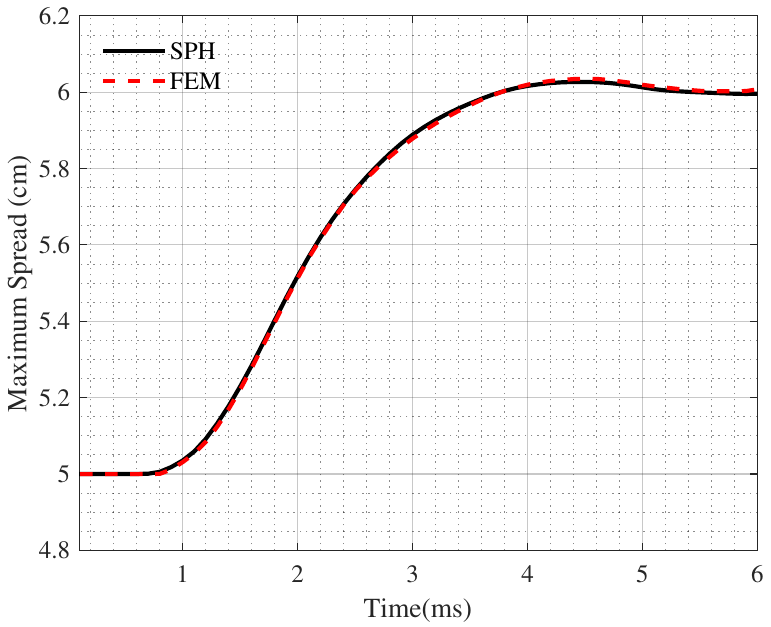}
	\caption{Comparision of maximum spread between SPH simulation and FEM model}
	\label{fig:max_spread_comparision}
\end{figure}

Deformed shapes of the soil cylinder along with the pressure contour at different time instants are shown in Figure \ref{fig:soil_drop_snapshot}. From the pressure contour, it is clear that initially, the surface cylinder contact is under compression, after a certain time (see Figure \ref{fig:soil_drop_snapshot}) the reflected shock wave is generated, which causes lateral spreading of the cylinder. Also, the spread of the contact surface is occurring at a rapid rate initially, and after sometimes, it reaches a constant value. Maximum spread width also follows the same trend.
For the validation purpose, a similar 2D plane strain problem is simulated via FEM in Abacus. Figure \ref{fig:max_spread_comparision} shows the comparison of the maximum spread of the soil cylinder between SPH and FEM simulations. The initial flat line indicates the state before the impact. As evident from Figure \ref{fig:max_spread_comparision} the time evolution of the maximum spread obtained from FEM and SPH are shown to have good agreement. 

\section{Slope failure analysis on cohesive soil} \label{SlopeFailure}
Large deformation and slope failure of a vertical pit of cohesive soil under the influence of gravity is performed in this section. Understanding the post-failure profile is important to ascertain the damage risk associated with landslides and/or related phenomena. Slope failure in cohesive soil involves the formation of a high plastic strain zone and shear band, which propagates as the failure initiates. SPH simulation of the similar problem is also attempted by a few other researchers \citep{bui2008lagrangian, islam2022large} to highlight the issues related to tensile instability. 

\begin{figure}[H]
	\centering
	\includegraphics[width=8cm]{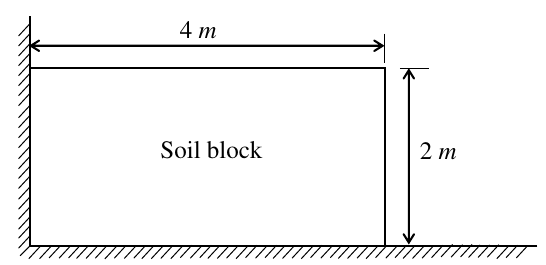}
	\caption{Initial configuration}
	\label{fig:block_fig_1}
\end{figure}

The initial geometry of the soil mass is shown in Figure \ref{fig:block_fig_1}. The soil domain is discretized by 13041 particles. For interaction between the soil mass and the rigid wall, no-slip boundary conditions \cite{morris1997modeling} as discussed in the previous section is implemented. The material properties and computational data are provided in Tables \ref{tab:slope_table_1} and \ref{tab:slope_table_2} respectively. 

\begin{table}[H]
	\centering 
	\caption{Material properties of soil}
	\setlength{\tabcolsep}{5pt}  
	\renewcommand{\arraystretch}{1.5}  
	\begin{tabular}{c  c  c  c  c }
		\hline
		$\rho_0$ &  $E$ & Poisson's ratio$(\nu)$ & Cohesion $(c)$ & Internal friction angle$(\phi$)\\
		\hline
		1850 kg/m$^3$ & 1.5 MPa & 0.2 & 5 kPa & $20^\circ$ 
	\end{tabular}
	\label{tab:slope_table_1}
\end{table}
\begin{table}[H]
	\centering 
	\caption{SPH computational data for simulating slope failure}
	\setlength{\tabcolsep}{5pt}  
	\renewcommand{\arraystretch}{1.5}  
	\begin{tabular}{c  c  c  c  c }
		\hline
		Particle spacing $(s)$ & Smoothing length $(h)$ & $\Delta t$  & \multicolumn{2}{c}{Artificial viscosity coefficients}\\
		\cline{4-5}
		&&& $\alpha$ &$\beta$ \\
		\hline
		0.025 m & 0.0375 m & $5\times10^{-5}$ s & 1.0 & 0
	\end{tabular}
	\label{tab:slope_table_2}
\end{table}

\begin{figure}[H]
	\centering
	\captionsetup[subfigure]{labelformat=empty}
	\begin{subfigure}[b]{0.75\textwidth}
		\centering
		\includegraphics[width=\textwidth]{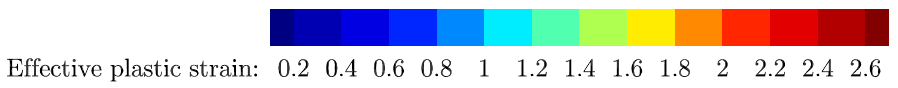}
		\label{fig:contour_snapshots}
	\end{subfigure}
	\begin{subfigure}{\textwidth}
		\centering
		\includegraphics[width=0.40\textwidth]{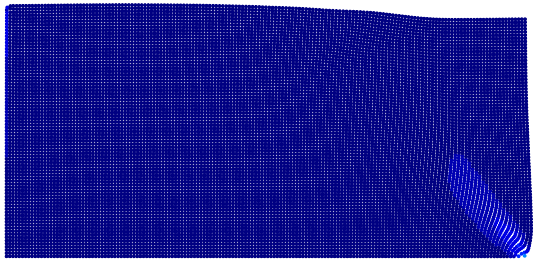}
		\hspace{1cm}
		\includegraphics[width=0.40\textwidth]{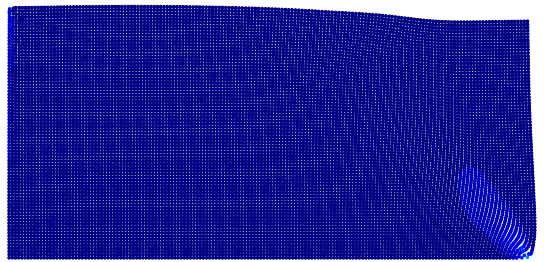}
		\caption{$t=0.3$ s}
		\label{fig:snapshot_0.3}
	\end{subfigure}
	\begin{subfigure}{\textwidth}
		\centering
		\includegraphics[width=0.40\textwidth]{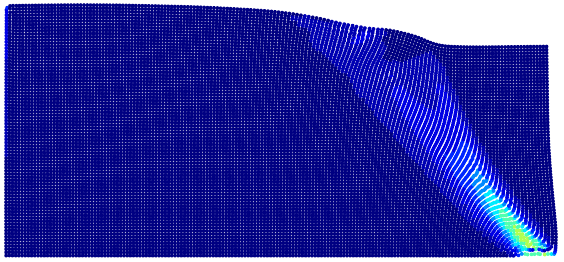}
		\hspace{1cm}
		\includegraphics[width=0.40\textwidth]{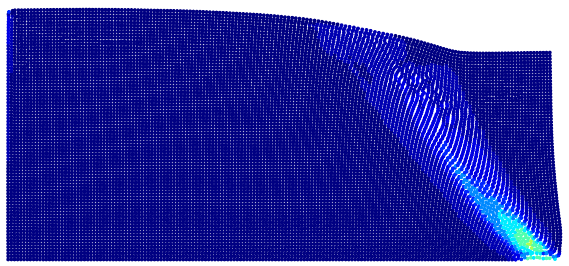}
		\caption{$t=0.6$ s}
		\label{fig:snapshot_0.6}
	\end{subfigure}
	\begin{subfigure}{\textwidth}
		\centering
		\includegraphics[width=0.40\textwidth]{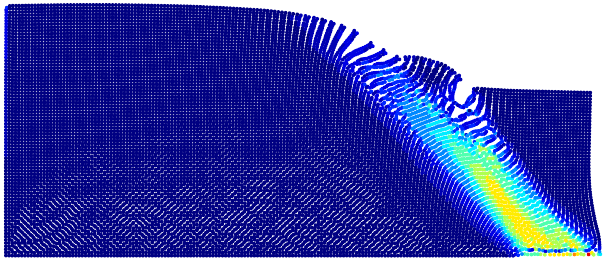}
		\hspace{1cm}
		\includegraphics[width=0.40\textwidth]{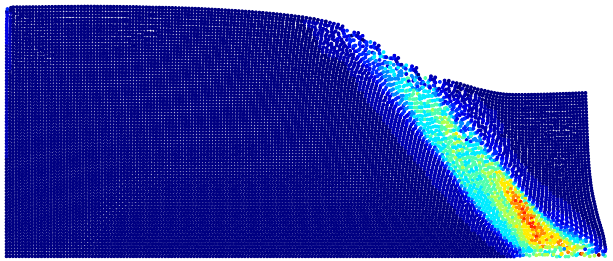}
		\caption{$t=1.2$ s}
		\label{fig:snapshot_1.2}
	\end{subfigure}
	\begin{subfigure}{\textwidth}
		\centering
		\includegraphics[width=0.40\textwidth]{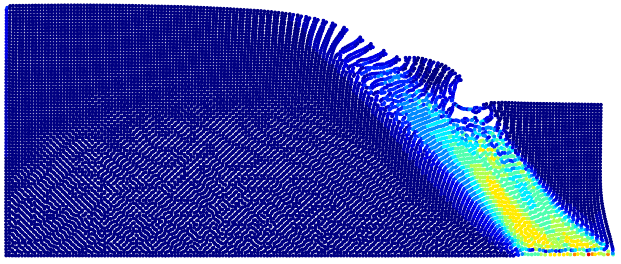}
		\hspace{1cm}
		\includegraphics[width=0.40\textwidth]{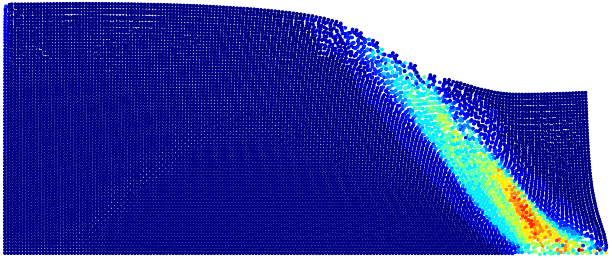}
		\caption{$t=2.0$ s}
		\label{fig:snapshot_2}
	\end{subfigure}
	\begin{subfigure}[b]{0.40\textwidth}
		\caption{(a)}
	\end{subfigure}
	\begin{subfigure}[b]{0.40\textwidth}
		\caption{(b)}
	\end{subfigure}
	\hfill
	\caption{Deformation pattern and effective plastic strain for different time instants: (a) Conventional SPH,(b) with Adaptive kernel approach}
	\label{fig:snapshots_coh}
\end{figure}

Deformation configurations of the soil block at different time instants obtained from the standard SPH and the proposed adaptive SPH are shown in Figure \ref{fig:snapshots_coh}. It may be observed that initially, a zone with high plastic deformation forms near the bottom corner, which then propagates towards the top surface and causes material slippage. While capturing the phenomena, the standard SPH suffers severe tensile instability, which manifests itself in the form of unphysical particle separations as well as unrealistic crack formations along the high plastic deformation zone (Figure \ref{fig:snapshots_coh}a). Another numerical instability may be seen near the bottom boundary, which propagates with time and may be ascribed to the zero energy oscillations (\cite{nguyen2017new} \cite{islam2022large}). The adaptive SPH, on the other hand, allows a stable computation (Figure \ref{fig:snapshots_coh}b). 

\begin{figure}[H]
	\centering
	\begin{subfigure}[b]{0.45\textwidth}
		\centering
		\includegraphics[width=\textwidth]{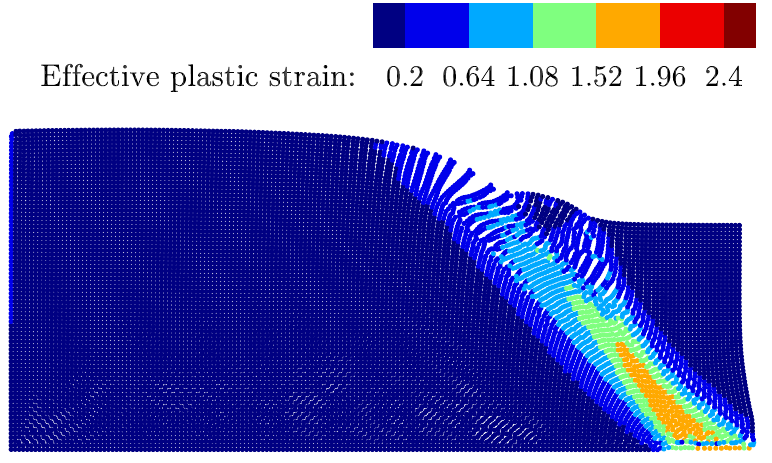}
		\caption{}
		\label{fig:conv_1s}
	\end{subfigure}
	~
	\begin{subfigure}[b]{0.45\textwidth}
		\centering
		\includegraphics[width=\textwidth]{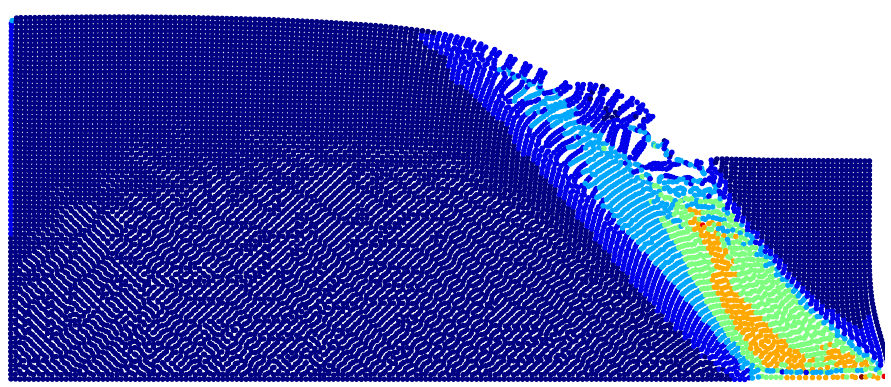}
		\caption{}
		\label{fig:art_0.1_1s}
	\end{subfigure}\\
	~
    \begin{subfigure}[b]{0.45\textwidth}
	    \centering
	    \includegraphics[width=\textwidth]{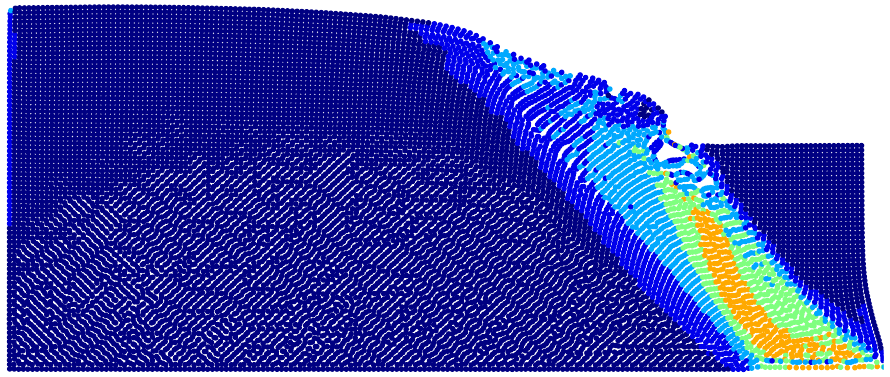}
	    \caption{}
	    \label{fig:art_0.2_1s}
    \end{subfigure}
	~
    \begin{subfigure}[b]{0.45\textwidth}
	    \centering
	    \includegraphics[width=\textwidth]{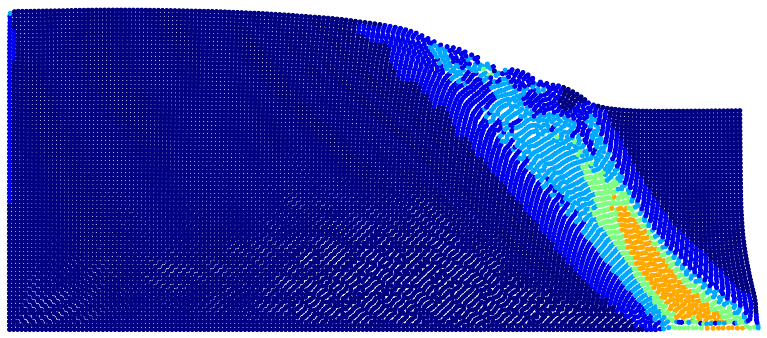}
	    \caption{}
	    \label{fig:art_0.3_1s}
    \end{subfigure}
	~	
	\begin{subfigure}[b]{0.45\textwidth}
		\centering
		\includegraphics[width=\textwidth]{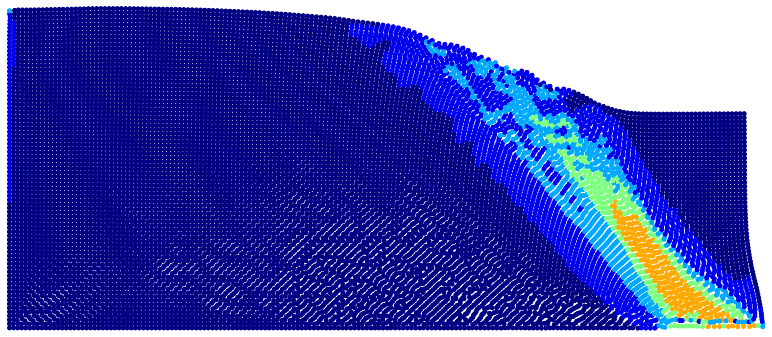}
		\caption{}
		\label{fig:art_0.5_1s}
	\end{subfigure}
	~
	\begin{subfigure}[b]{0.45\textwidth}
		\centering
		\includegraphics[width=\textwidth]{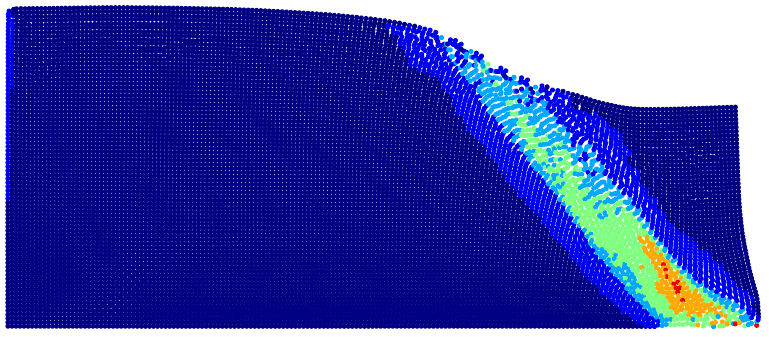}
		\caption{}
		\label{fig:adap_1s}
	\end{subfigure}
	\hfill
	\caption{Deformation pattern along with effective plastic strain: (a) Conventional SPH, (b) with artificial stress coefficient $\epsilon=0.1$, (c) with $\epsilon=0.2$, (d) with $\epsilon=0.3$, (e) with $\epsilon=0.5$, (f) with Adaptive kernel approach}
	\label{fig:cohesive_diff_scheme_1s}
\end{figure}

\begin{figure}[H]
	\centering
	\captionsetup[subfigure]{labelformat=empty}
	\begin{subfigure}[b]{0.75\textwidth}
		\centering
		\includegraphics[width=\textwidth]{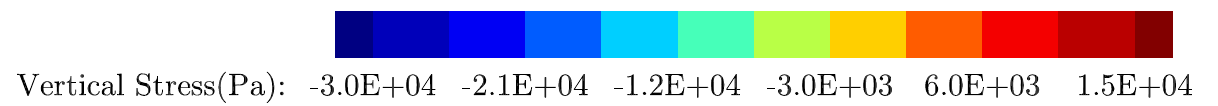}
		\label{fig:contour_1}
	\end{subfigure}
	\begin{subfigure}{\textwidth}
		\includegraphics[width=0.33\textwidth]{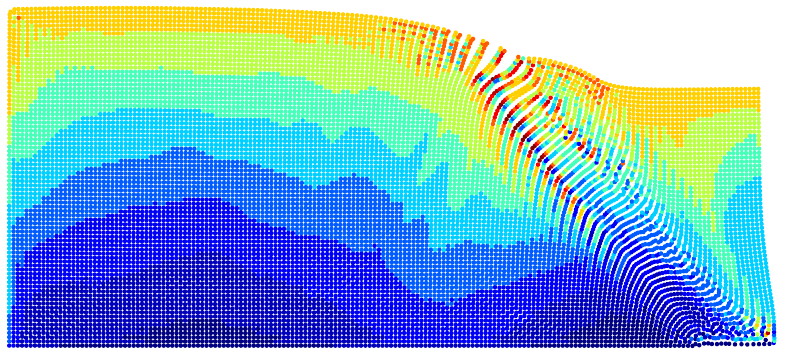}
		\hspace{0cm}
		\includegraphics[width=0.32\textwidth]{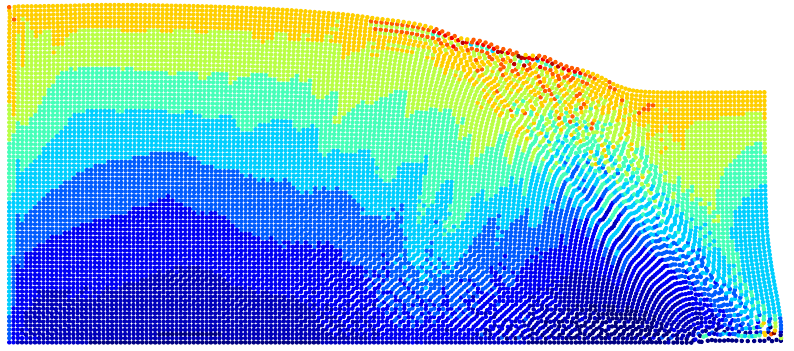}
		\hspace{0cm}
		\includegraphics[width=0.32\textwidth]{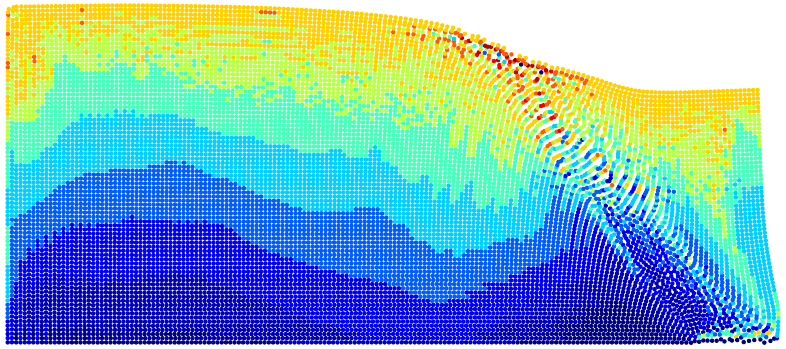}
		\caption{$t=0.8$ s}
		\label{fig:ver_str_0.8}
	\end{subfigure}
	\begin{subfigure}{\textwidth}
		\includegraphics[width=0.33\textwidth]{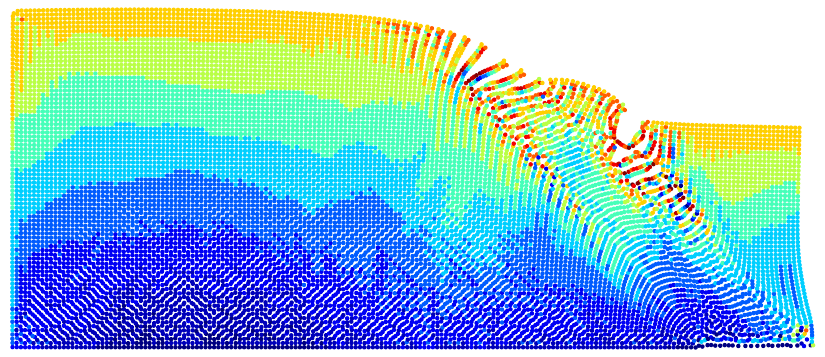}
		\hspace{0cm}
		\includegraphics[width=0.32\textwidth]{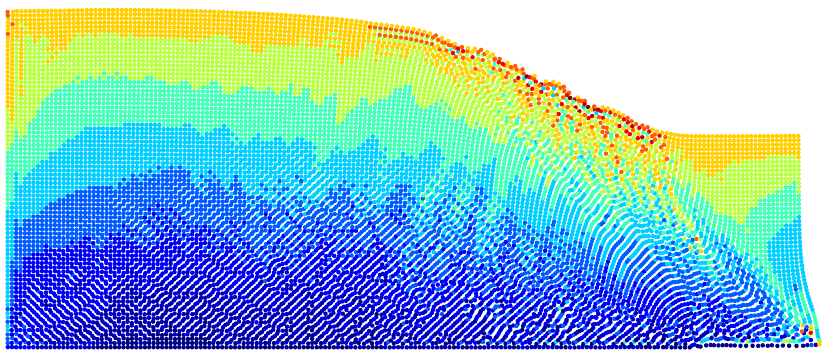}
		\hspace{0cm}
		\includegraphics[width=0.32\textwidth]{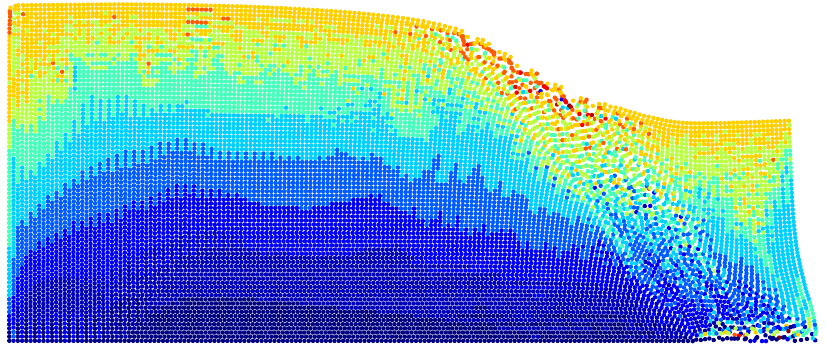}
		\caption{$t=1.2$ s}
		\label{fig:ver_str_1.2}
	\end{subfigure}
	\begin{subfigure}{\textwidth}
		\includegraphics[width=0.33\textwidth]{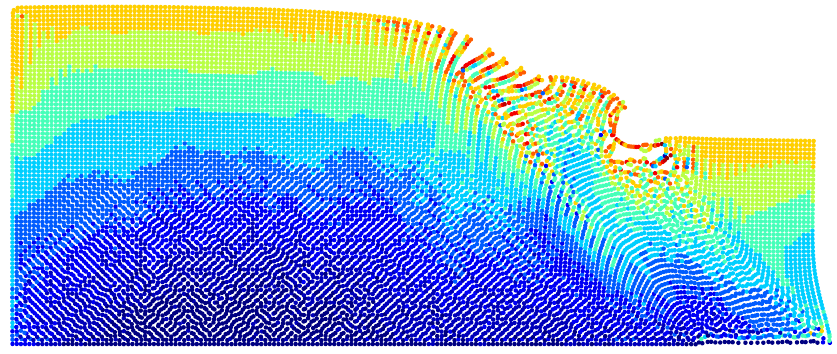}
		\hspace{0cm}
		\includegraphics[width=0.32\textwidth]{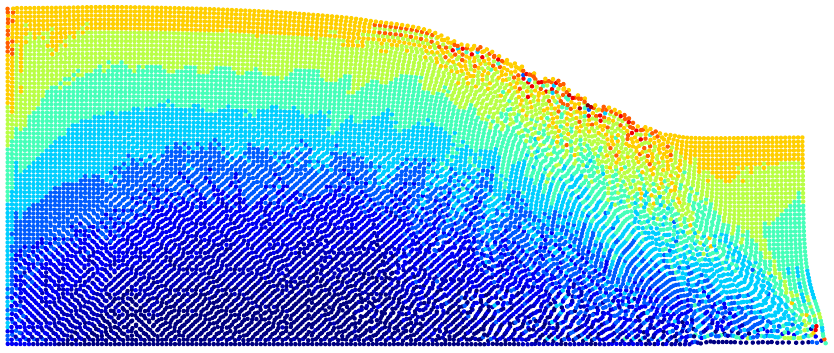}
		\hspace{0cm}
		\includegraphics[width=0.32\textwidth]{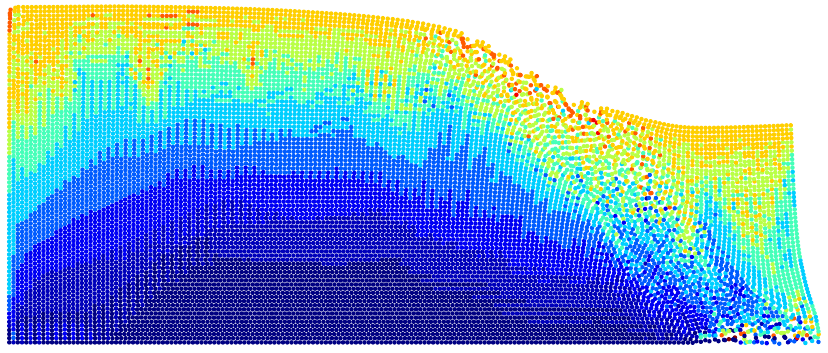}
		\caption{$t=1.6$ s}
		\label{fig:ver_str_1.6}
	\end{subfigure}
	\begin{subfigure}[b]{0.34\textwidth}
		\caption{(a)}
	\end{subfigure}
	\begin{subfigure}[b]{0.32\textwidth}
		\caption{(b)}
	\end{subfigure}
	\begin{subfigure}[b]{0.32\textwidth}
		\caption{(c)}
	\end{subfigure}
	\hfill
	\caption{Variation of vertical stress $(\sigma_{yy})$ for different time instants: (a) Conventional SPH, (b) with artificial stress coefficient $\epsilon=0.5$, (c) with Adaptive kernel approach}
	\label{fig:ver_str_coh}
\end{figure}

The artificial stress \citep{monaghan2000sph, gray2001sph} is a widely used technique to deal with tensile instability in SPH. Therein, an artificial stress term is added in the momentum equation, and the strength of the artificial term is regulated by a user defined parameter $\epsilon$. In order to see how the proposed methodology stands vis-a-vis SPH with artificial stress term, simulations are performed with different values of $\epsilon$. The deformed configurations at $t=2$ s obtained with $\epsilon = 0.1$, $\epsilon = 0.2$ and $\epsilon = 0.3$ are compared with that obtained via the proposed adoptive algorithm in Figure \ref{fig:cohesive_diff_scheme_1s}. It is observed that $\epsilon < 0.3$ fails to suppress the tensile instability. The artificial stress coefficient as $0.3$ was also suggested by \citet{monaghan2000sph} for simulating solid mechanics problems. \citet{bui2008lagrangian} suggested a value of $\epsilon=0.5$ for simulating geomaterials. Although the artificial stress approach with $\epsilon \geq 0.3$ can suppress the tensile instability effectively, the choice of the value of required $\epsilon$ for a given problems still remains a computational issue. Moreover, the use of artificial stress can not treat the instability caused by spurious zero energy oscillations of particles and causes irregular stress distribution (Figure \ref{fig:ver_str_coh}). From Figure \ref{fig:adap_1s}, it is clearly evident that the Adaptive algorithm based on the pressure zone approach not only removes the tensile instability that arises at the high plastic deformation region but also completely avoids the formation of any kind of particle disturbances near wall boundary caused by zero energy modes.

Particle configurations along with vertical stress $(\sigma_{yy})$ contour are plotted for different time instants $(t=0.8, 1.2, 1.6$ s) in Figure \ref{fig:ver_str_coh}. Initially, the difference between stress distributions is negligible. But as time progresses $(t=1.2$ s and $t=1.6$ s) the irregularity of particles near the bottom boundary becomes prominent and it causes irregular stress distribution corresponding to conventional SPH as well as with artificial stress term. But stress contour remains smooth, and particle configurations remain regular, corresponding to SPH using adaptive kernel and pressure zone approach. The disturbances near the bottom boundary keep on increasing with time, although the final deformed pattern of the soil slope becomes completely at rest. 

\section {Conclusion} \label{Conclusion}
In the present study, a stable elastic-plastic SPH framework incorporating Drucker-Prager yield criteria is developed for large deformation analysis of geomaterials. In order to address the issue of tensile instability, a pressure zone-based adaptive approach is proposed. Herein, the shape of the kernel function is continuously modified depending on the state of stress such that the inter particle interaction always remains stable. A B-spline basis defined over a symmetric but variable knot vector is used as the kernel function. The shape of the kernel is then modified by changing the positions of the intermediate knots. For a ready reference, a flow chart outlining the different steps in the algorithm is also provided. Finally, the developed algorithm is applied to simulate the large deformation and slope collapse of a cohesive soil pit. In the process, the efficacy of the algorithm is demonstrated. The proposed algorithm also treats any possibility of the formation of spurious zero energy oscillations, which remains untreated when using the artificial stress method. The numerical predictions via the proposed algorithm are assessed in comparison with those obtained using the standard SPH and the artificial stress based SPH with different values of artificial stress coefficients. Tensile instability problem is dealt with more effectively using the adaptive kernel approach, as seen by the comparison.

\bibliographystyle{spbasic_unsort}
\bibliography{Paper_Soil}

\end{document}